\documentclass[11pt]{article}

\usepackage{authblk}
\usepackage[margin=1in]{geometry}
\usepackage{amssymb,amsmath,amsthm,amsfonts,thmtools}
\usepackage{color}
\usepackage[export]{adjustbox}
\usepackage{yfonts}
\usepackage{mathrsfs}

\usepackage{xspace}

\usepackage{verbatim}

\usepackage{graphicx,lipsum}
\usepackage[font=small]{caption}
\usepackage{subcaption}
\usepackage[utf8]{inputenc}

\usepackage{url}
\usepackage[all]{xy}
\usepackage{rotating}
\usepackage{ifpdf}
\usepackage{tcolorbox}
\usepackage{lipsum}

\usepackage{mdframed}             
\usepackage{xifthen} 
\usepackage{mathtools}  

\DeclareMathAlphabet\mathbfcal{OMS}{cmsy}{b}{n}

\usepackage{comment}
\usepackage[T1]{fontenc}

\usepackage{array}
\usepackage{enumitem}

\newcommand{\calT}{\ensuremath{\mathcal T}\xspace}

\newcommand{\tildeO}{{\tilde{O}}}

\colorlet{darkgreen}{green!45!black}

\newcommand{\ignore}[1]{}

\newcommand{\margincomment}[2]%
{\marginpar{\footnotesize\raggedright {\color{red}#1}: #2}}

\newcommand{\myparagraph}[1]{{\medskip\noindent\textbf{#1}}}

\newcommand{\braced}[1]{{ \left\{ {#1} \right\} }}

\newcommand{\reals}{{\mathbb R}}
\newcommand{\integers}{{\mathbb Z}}

\newcommand{\edgeto}{{\mspace{-0.01mu}\rightarrow\mspace{-0.01mu}}}

\newcommand{\weight}{\ell}
\newcommand{\distance}{\delta}

\newcommand{\half}{\textstyle{\frac{1}{2}}}

\newcommand{\ListLengths}{\setlength{\itemsep}{0ex}\setlength{\topsep}{1ex}\setlength{\partopsep}{0ex}}

\graphicspath{{./}} 

\title{A Refutation of Elmasry's $\tildeO(m\sqrt{n})$-Time Algorithm for Single-Source Shortest Paths\thanks{Research supported by NSF grant CCF-2153723.}}

\author[1]{Sunny Atalig}
\author[1]{Marek Chrobak}

\affil[1]{University of California at Riverside, USA}

\begin{document}

\maketitle

\begin{abstract}
In this note we examine the recent paper ``Breaking the Bellman-Ford Shortest-Path Bound''
by Amr Elmasry, where he presents an algorithm for the single-source shortest path
problem and claims that its running time complexity is $\tildeO(m\sqrt{n})$,
where $n$ is the number of vertices and $m$ is the number of edges.
We show that his analysis is incorrect, by providing an example of a weighted graph
on which the running time of his algorithm is $\Omega(mn)$.
\end{abstract}


\section{Introduction}
\label{sec: introduction}


\myparagraph{The SSSP problem.}
In the single-source shortest path problem (SSSP), we are given a graph $G$ with weights on edges and
a designated source vertex $s$, and the goal is to compute the shortest paths in $G$ from $s$ to all
other vertices. For simplicity, throughout this paper we assume that $G$ does not have any negative
cycles, so that the shortest paths are well defined.
We use notations $n$ and $m$ for the number of vertices and edges of $G$, respectively.

SSSP is a classical and extensively studied problem in the theory of combinatorial algorithms.
The Bellman-Ford algorithm~\cite{1955_shimbel_structure_in_communication_nets,1958_bellman_routing_problem,1959_moore_shortest_path,1956_ford_network_flow_theory}
 solves the SSSP problem in time $O(mn)$. In 2023, a major breakthrough was obtained by
Fineman~\cite{2024_Fineman_single-source_shortest_paths}, who provided
a randomized algorithm with running time $\tildeO(m n^{8/9})$.
Subsequently,  Huang, Jin and Quanrud~\cite{2024_Huang_etal_faster_shortest_paths_hop_distance,2025_Huang_etal_faster_shortest_paths_hop_distance}
improved the running time to $\tildeO(mn^{4/5})$,
and even more recently~\cite{2025_huang_etal_faster_shortest_paths_bootstrapping}
claimed an additional improvement to $\tildeO(mn^{3/4} + m^{4/5}n)$.

In an independent work, Elmasry~\cite{2024_elmasry_breakingbellmanford} developed a very different algorithm,
claiming that its running time is $\tildeO(m\sqrt{n})$. If valid, this result would constitute a substantial
improvement over the work in~\cite{2024_Fineman_single-source_shortest_paths,2024_Huang_etal_faster_shortest_paths_hop_distance,2025_huang_etal_faster_shortest_paths_bootstrapping},
especially that his algorithm is simple and deterministic.


\myparagraph{Notation and terminology.}
By $G = (V,E)$ we will denote the input graph, and the weight function is $\weight \colon E \to \integers$.
We write $\weight(e)$ or $\weight_e$ for the weight of an edge $e$. If $e = u\edgeto v$, we denote this weight
by $\weight(u,v)$ or $\weight_{u,v}$.
{
The length of a path is given by summing the weights of its edges,
and the distance $\distance(u,v)$ denotes the minimum length of a path from $u$ to $v$.
(We permit the empty path, and the distance is $\infty$ if no path from $u$ to $v$ exists.)
We extend this notation to sets, letting $\distance(X,v) = \min_{u \in X} \distance(u,v)$. 
}
 
By a \emph{potential function} we mean an arbitrary function $\phi : V\to \reals$.
The potential value on $v$ will be denoted by $\phi_v$ or $\phi(v)$.

Given any weight function $\weight$ and a potential $\phi$, by $\weight_\phi$ we define
the $\phi$-adjusted weight function $\weight_\phi(u,v) = \weight (u,v) + \phi(u) -\phi(v)$.
Potential $\phi$ is called \emph{valid} for $\weight$ if $\weight(e) \ge 0$ implies $\weight_\phi(e)\ge 0$; that is
non-negative edges remain non-negative after the adjustment.
The weight function $\weight_\phi$ has the same shortest-path structure as $\weight$,
namely for any two paths $P$ and $Q$ that have the same endpoints we have
$\weight_\phi(Q) - \weight_\phi(P) = \weight(Q) - \weight(P)$. So, although the path lengths may
change by the adjustments, their length difference remains the same.
For consistency with~\cite{2024_elmasry_breakingbellmanford}, we will refer
to $\weight_\phi$ as the \emph{reduced weight}.


\myparagraph{Neutralizing negative weights.}
The overall goal of SSSP algorithms, if negative weights are allowed, is to find a  \emph{neutralizing} potential function,
namely a function $\phi$ for which $\weight_\phi$ is non-negative. 
This does not change shortest-paths from $s$, so
then one can run Dijkstra's algorithm on $\weight_\phi$ to find the shortest-paths.
This is the underlying idea behind many shortest-path algorithms
and is often attributed to Johnson~\cite{1997_Johnson_efficient_shortest_paths} (although Johnson himself
attributes it to some earlier sources, including~\cite{1972_edmonds_karp_network_flow_problems}).

For any non-empty set $X\subseteq V$, consider the potential function $\phi(v) = \distance(X,v)$, for all $v\in V$.
(We assume here that all vertices are reachable from $X$.)
Then, trivially, for each edge $u\edgeto v$ we have $\phi(v) \le \phi(u) + \weight(u,v)$, so 
$\phi$ is a neutralizing potential. In particular, when $X = V$, this potential will be called the \emph{Johnson potential}.


\myparagraph{Elmasry's algorithm.}
The algorithm proposed by Elmasry in~\cite{2024_elmasry_breakingbellmanford} is in essence an alternative way
of computing Johnson's potential. It computes this potential incrementally, at each iteration
computing a new increment, which is by itself also a potential. The sum of the potentials from all iterations will neutralize all negative edges.

We refer to the potential from each iteration as \emph{Elmasry's potential}, and it is defined as follows:
Let $G$ be the current graph, and
 let an \emph{nbp-path in $G$} (negative-before-positive) be any path on which
 all negative edges precede all positive edges. (In particular, such paths may not have any
 negative edges or any positive edges, or can even be empty.)
 For each $v$, we let $\eta(v)$ to be the minimum length of an nbp-path in $G$ starting
 at any vertex and ending at $v$. (By definition, this path will be either empty or start
 with a negative edge.)

Observe that Elmasry's potential is valid. To see this, consider any edge $u\edgeto v$
with non-negative weight, $\weight(u,v)\ge 0$. Let $P$ be the 
nbp-path that realizes the potential for $u$. Then, since $\weight(u,v)\ge 0$,
$Pv$ is also an nbp-path, and thus 
$\eta(u) + \weight(u,v) =  \weight(P) + \weight(u,v) = \weight(Pv) \ge \eta(v)$,
by the definition of $\eta$. So $\weight_\eta(u,v) = \weight(u,v) + \eta(u) - \eta(v) \ge 0$.

Elmasry's algorithm consists of a sequence of iterations. 
In each iteration, denoting by $G$ the current graph before the iteration, the algorithm
computes Elmasry's potential $\eta$ for $G$, replaces the weights by the reduced weights, and continues on the new graph.
This potential $\eta$ can be computed in nearly linear time as follows:
Initialize $\eta(u) = 0$ for all $u$.
Let $G^-$ be the subgraph of $G$ induced by non-positive edges.
We can assume that $G^-$ is acyclic. (If $G$ has a negative cycle, we can
stop. If $G$ has a $0$-weight cycle, the vertices in this cycle can be
contracted into one vertex.) First, propagate $\eta$ in $G^-$, following the
topological ordering of $G^-$. Then, propagate $\eta$ in $G^+$ (the graph
induced by non-negative edges) using Dijkstra's algorithm.

{
Superficially, Elmasry's algorithm looks similar to ``Bellman-Ford-Dijkstra'' (BFD) hybrid variants,
which essentially compute the shortest paths by alternating between processing negative edges and non-negative edges
(see \cite{2017_dinitz_itzhak_hybrid_bellman_ford_dijkstra,2022_bernstein_etal_negative_weights,2024_Fineman_single-source_shortest_paths}).
While BFD can be used to efficiently compute shortest paths with a bounded number of negative edges,
one can easily find graphs where the algorithm runs in $\Omega(mn)$ time when used directly to compute the Johnson potential.
(For example, take $G$ to be a path which alternates between negative and positive edges.)
Elmasry's algorithm is non-trivial because zero-weight edges may appear anywhere in an nbp-path;
one can interpret his algorithm as instead alternating between processing \emph{non-positive} edges and non-negative edges.
This difference makes the algorithm resistant to the simple counterexamples for BFD.
}
%


 \myparagraph{Snakes and outline of analysis.}
A key concept in Elmasry's analysis is that of ``snakes''.
A \emph{snake} is defined to be an inclusion-maximal\footnote{In~\cite{2024_elmasry_breakingbellmanford},
the definition (top of page~5) actually says ``maximal-length'', but, based on its later use, it's clear that 
the author intended it to mean ``inclusion-maximal''. Any ambiguity related to this definition does not affect our counter-example.}
path consisting of zero-weight edges followed by exactly one negative edge, 
which we call the snake's \emph{head}. The rest of the snake (consisting of zero-weight edges) is referred to as \emph{tail}.
This definition is based on the current set of weights after previous adjustments, 
so obviously the set of snakes can change in each iteration.  Because Elmasry's potential is valid, no ``new'' snake heads 
can appear in the reduced weight function, 
after applying the adjustment (i.e.\ a head will always correspond to a negative edge in the original weights $\ell$), 
but their tails may change, 
and snakes may possibly disappear (i.e.\ when a negative head edge gets neutralized, the associated snakes from the previous iteration are no longer snakes).

The overall idea of the analysis has two key ingredients:
\begin{itemize}

\item First, show that after $j-1$ iterations each snake will have length at least $j$ (Lemma~6 in~\cite{2024_elmasry_breakingbellmanford}).

\item Next, show that in each iteration the first snake in the
	shortest-path tree $\calT$~\footnote{This is the shortest-path tree that
	realizes the Johnson potential. Technically, it's a forest, but it can
	be viewed as a tree if a virtual node $s^\ast$ is added with $0$-weight edges to the nodes in $G$.}
	either disappears or joins the next snake (the proof of Theorem~1 in~\cite{2024_elmasry_breakingbellmanford}).

\end{itemize}

The second statement implies that the first snake grows by at least $j$ in iteration $j$, so its growth in the sequence
of iterations would actually be quadratic, implying a bound of $\sqrt{n}$ on the number of iterations.


\myparagraph{Flaws in~\cite{2024_elmasry_breakingbellmanford}.}
We show that the analysis in~\cite{2024_elmasry_breakingbellmanford} is incorrect. First, we address the
proof of Lemma~6 in~\cite{2024_elmasry_breakingbellmanford}. This is an inductive argument, claiming that
in each step the minimum length of snakes increases by at least $1$. 
In Section~\ref{sec: counterexample for growing snakes lemma} we show an example where
this is not the case.

Building on this example, in Section~\ref{sec: lower bound for directed acyclic graphs}
we construct a graph on which the algorithm from~\cite{2024_elmasry_breakingbellmanford}
makes $\Omega(n)$ iterations. As each iteration runs in time $\Omega(m)$, this shows that that the worst-case running time
of this algorithm is $\Omega(mn)$ on some graphs, refuting the main result in~\cite{2024_elmasry_breakingbellmanford}.

For the sake of completeness,
we have also analyzed the complexity of this algorithm when $G$ is just a path starting at the source vertex $s$.
In Section~\ref{sec: analysis for paths} we prove that on paths 
Elmasry's algorithm makes $O(\log n)$ iterations, and that this bound is optimal for paths.
While this is not of much interest of its own (because for acyclic graphs the SSSP problem
can be solved in linear time), we hope that this observation may be useful
in designing better algorithms for the general case.


\section{Counter-Example for the Growing Snakes Lemma}
\label{sec: counterexample for growing snakes lemma}
\begin{figure}[ht]
\begin{center}
	\includegraphics[width=5.5in]{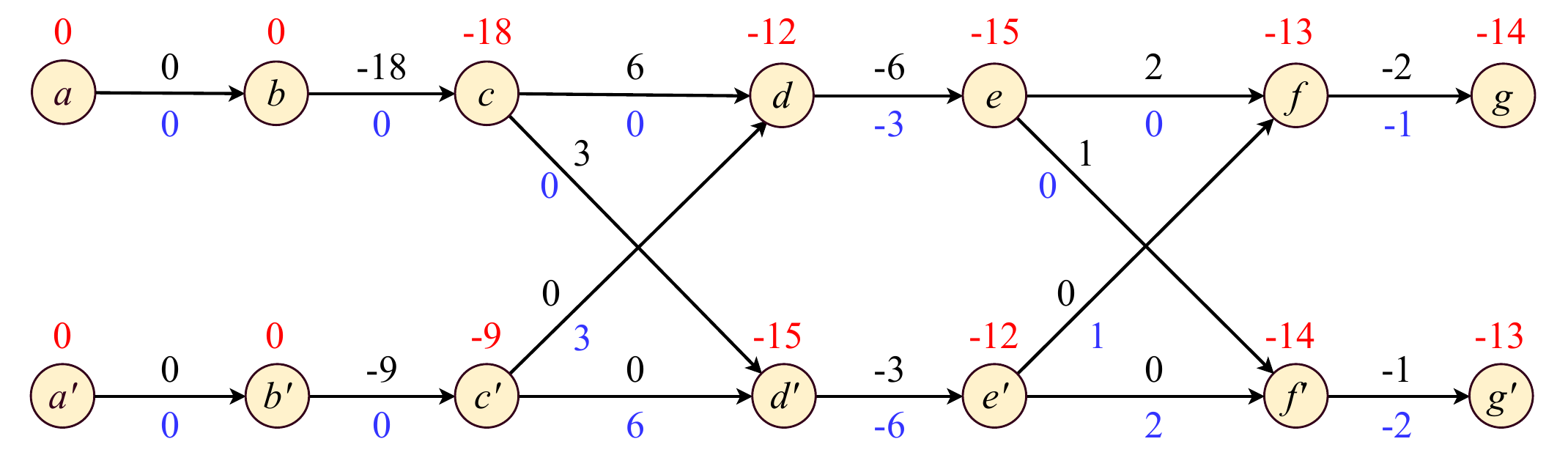}
	\caption{Example illustrating the flaw in the proof of Lemma~6.}
	\label{fig: counterexample lemma6}
\end{center}
\end{figure}

We start by showing that the argument in the proof of Lemma~6 in~\cite{2024_elmasry_breakingbellmanford} is flawed.
The main claim in this proof is that in each step the minimum snake length increases by at least $1$. The example
in Figure~\ref{fig: counterexample lemma6} shows that this is not always true.

Figure~\ref{fig: counterexample lemma6} shows the change resulting from one iteration.
The edge weights before the iteration are above the edges (in black). The values of Elmasry's potential $\eta$ are shown above the vertices (in red).
Recall that these values are computed in two sub-phases, by first propagating its values in $G^-$ and later in $G^+$.
For example, after the first sub-phase, the potential of vertex $d$ will be obtained from the path $a'\edgeto b' \edgeto c'\edgeto d$ in $G^-$, giving the
value $-9$, but in the second sub-phase it will be decreased, using edge $c\edgeto d$ in $G^+$, giving $\eta(d) = -12$.

Before the iteration, $G$ has $6$ snakes, all of length $2$:
\begin{equation*}
\begin{array}{ccccc}
a\edgeto b \edgeto c  &\quad& a'\edgeto b'\edgeto c' &\quad& c' \edgeto d \edgeto e
\\
c' \edgeto d'\edgeto e' && e'\edgeto f\edgeto g && e'\edgeto f'\edgeto g'
\end{array}
\end{equation*}

The reduced edge weights after the iterations are shown below the edges (in blue). Note that
there is still one snake of length $2$, namely $e \edgeto f'\edgeto g'$, contradicting the proof of  Lemma~6 in~\cite{2024_elmasry_breakingbellmanford}.

The reason why the snake $e'\edgeto f'\edgeto g'$ does not grow (is not replaced by a longer snake ending at $g'$) can be explained as follows:
While the the shortest nbp-path ending in $g'$ is 
$a'\edgeto b'\edgeto c'\edgeto d'\edgeto e'\edgeto f'\edgeto g'$ (of length $-13$), 
the shortest nbp-path ending at $f'$  is
$a'\edgeto b' \edgeto c' \edgeto d \edgeto e \edgeto f'$ (of length $-14$). That is, the nbp-path for $g'$ is not an extension of the nbp-path for $f'$ (which has
negative length), and because of this, the edge $f'\edgeto g'$ will remain negative.
The intuition is, the positive-weight edge $e\edgeto f'$ ``blocks'' the update along edge $f'\edgeto g'$.
For similar reasons, edge $d\edgeto e$ will remain negative, so the newly created snake $e\edgeto f'\edgeto g'$
will only have length $2$.


\section{Lower Bound for Directed Acyclic Graphs}
\label{sec: lower bound for directed acyclic graphs}

Here we construct a DAG where Elmasry's algorithm executes $\Omega(n)$ iterations.
The construction is based on the example from Figure~\ref{fig: counterexample lemma6}. As illustrated by that example,
relatively light positive edges may block large negative weights from being propagated along the paths in $G$, and thus from
neutralizing some negative edges, preventing some snakes from growing.
Another observation is that starting at $c$ and $c'$, the weights between the top
and bottom portions of the graph simply flip vertically, essentially not changing this part of the graph.
We can therefore chain these ``gadgets'', 
so that the positive edges prevent the Elmasry potential from neutralizing most negative edges, 
while at the same time ensuring that there remain blocking positive edges in the next phase. This is done using exponentially large weights.

\medskip

Some general observations first.
The $O(\sqrt{n})$-iterations claim in~\cite{2024_elmasry_breakingbellmanford}
ultimately rests on the claim that after applying Elmasry's potential $\eta$, every negative edge either gets neutralized, 
or a corresponding snake gets ``joined'' by the snake whose head (by definition, negative) gets neutralized. What is definitely true is the following: 
\emph{if a negative edge $u\edgeto v$ is not neutralized, then the path associated with $\eta(u)$ must have negative total weight and contain a 
negative edge followed (not necessarily directly) by a positive edge}. 
More precisely, we can then say that the path that realizes $\eta(u)$ has the form 
$x \rightsquigarrow y \edgeto z \rightsquigarrow p \edgeto q \rightsquigarrow u$, 
where edge $y\edgeto z$ is negative, edge $p\edgeto q$ is (strictly) positive, 
the paths $x \rightsquigarrow y$ and $q \rightsquigarrow u$ have only zero-weight edges,
on the path $z \rightsquigarrow p$ all negative edges precede all positive edges,
and the total path length $\eta(u)$ is negative.
Under these conditions, in the reduced weight function
the edges along the $x$-to-$u$ path will attain zero reduced weight, hence creating a snake with head $u\edgeto v$.
If $x \rightsquigarrow y \edgeto z$ and $q \rightsquigarrow u \edgeto v$ are both snakes then one can think of
the new snake  $x \rightsquigarrow y \edgeto z \rightsquigarrow p \edgeto q \rightsquigarrow u$ as resulting from
``joining'' $x \rightsquigarrow y \edgeto z$ with $q \rightsquigarrow u \edgeto v$.

There are at least two problems with this analysis. The first is that the sub-path $q \rightsquigarrow u \edgeto v$ 
is not necessarily a snake (i.e.\ it is possible for $q$ to have an incoming zero-weight edge), 
and so Elmasry's potential will only join a suffix of the snake ending at $u$, not the whole snake. 
The second issue is that the negative edge $y \edgeto z$ may not even get neutralized.
Both of these issues are already illustrated in the example from Figure~\ref{fig: counterexample lemma6}.

\medskip

\begin{figure}[ht]
\begin{center}
	\includegraphics[width=3.5in]{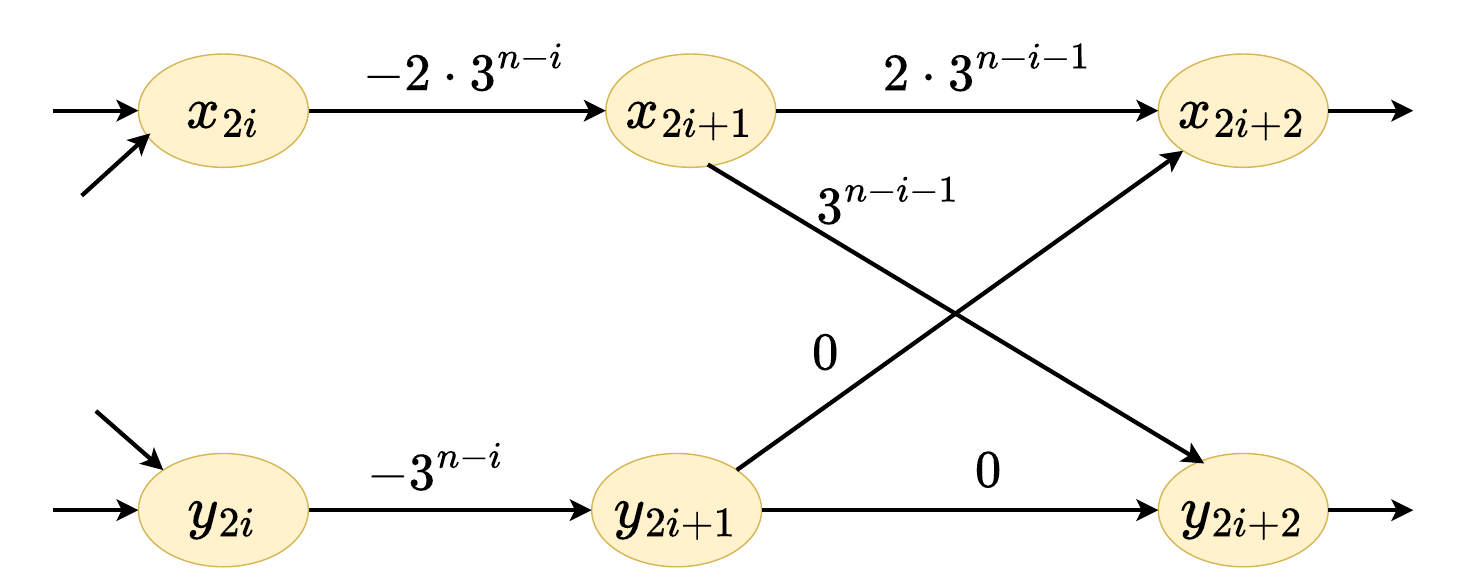}
	\caption{Edge weights in $G_n$.}
	\label{fig: edge weights in Gn}
\end{center}
\end{figure}

\begin{figure}[ht]
\begin{center}
	\includegraphics[width=5in]{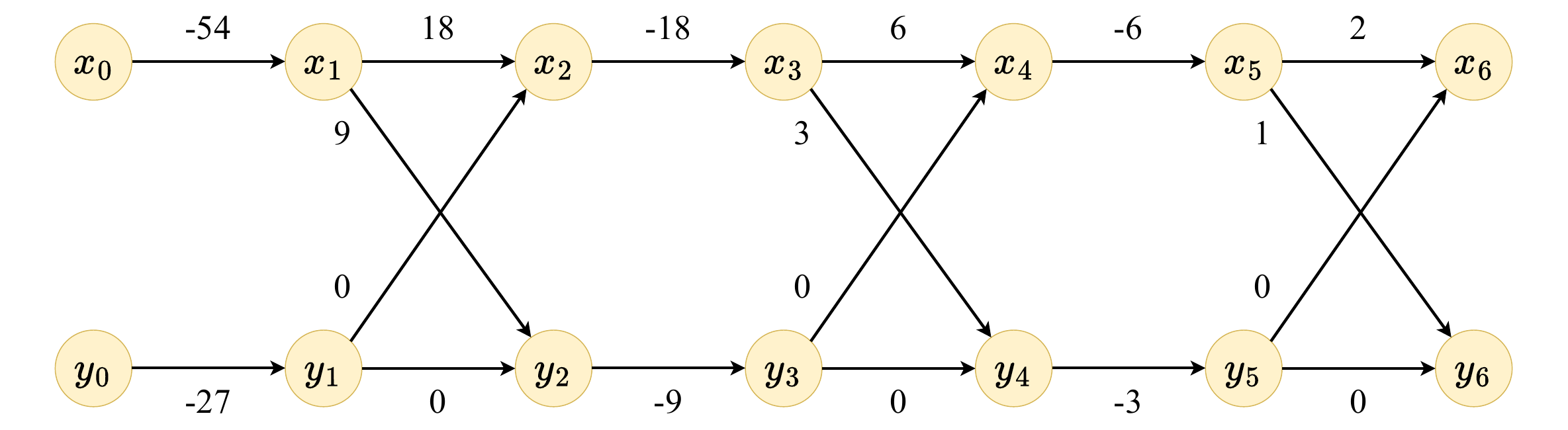}
	\caption{Graph $G_3$.}\label{fig: graph G3}
\end{center}
\end{figure}

We now give the family of graphs $G_n = (V_n,E_n)$ representing our counter-example. 
Denote by  $V_n$ the vertices $x_0, \ldots, x_{2n}, y_0 , \ldots y_{2n}$. 
To define $E_n$, for each $0 \le i < n$ add the following edges (see Figure~\ref{fig: edge weights in Gn}):
\begin{itemize}
\item  $x_{2i}\edgeto x_{2i+1}$, $x_{2i+1} \edgeto x_{2i+2}$ with weights $\ell(x_{2i}, x_{2i+1})= -2 \cdot 3^{n-i} $ and $\ell(x_{2i+1}, x_{2i+2}) = 2 \cdot 3^{n-i-1}$.
\item  $y_{2i} \edgeto y_{2i+1}$, $y_{2i+1} \edgeto y_{2i+2}$ with weights $\ell(y_{2i}, y_{2i+1})=  - 3^{n-i} $ and $\ell(y_{2i+1}, y_{2i+2}) = 0$.
\item  $x_{2i+1}\edgeto y_{2i+2}$, $y_{2i+1}\edgeto x_{2i+2}$ with weights $\ell(x_{2i+1}, y_{2i+2}) = 3^{n-i-1}$ and $\ell(y_{2i+1}, x_{2i+2}) = 0$.
\end{itemize}
Clearly we have $|V_n| = 4n+2$ and $|E_n| = 6n$. Graph $G_3$ is shown in Figure~\ref{fig: graph G3}.

For the analysis, it's useful to refer to the potential $\eta^{-}(v)$, which we define to be the Johnson potential on $G^{-}_n$. 
Recall that the Elmasry potential $\eta$ results from first computing $\eta^{-}$ and then propagating it through Dijkstra's algorithm. 
By induction on $n$, we show that $G_n$ takes at least $n$ iterations before all edges are neutralized. Trivially $G_1$ takes at least 1 iteration.

For the inductive step, we observe what happens after the first iteration, applying the Elmasry potential based on the weights defined above. 
First we determine $\eta^{-}$: by inspection, the sub-graph $G^{-}_n$ forms a directed tree on $V_n \setminus \braced{x_0, x_1}$ rooted at $y_0$,
 along with the disconnected edge $x_0\edgeto x_1$. Furthermore, the shortest path in $G^{-}_n$ ending at $v\in V_n \setminus \braced{x_0, x_1}$ 
 is given by the unique path from $y_0$. More precisely, the path lengths and potentials after the first sub-phase are
\begin{align*}
	\eta^{-} (y_{2i+1}) &= \eta^{-} (y_{2i+2}) = \textstyle -\sum_{j=0}^{i} 3^{n-j} 
			= \half (\, 3^{n-i}  - 3^{n+1} \,)
			 &&\qquad \text{ for $0\le i <n$} \\
	\eta^{-} (x_{2i}) &=  \eta^{-} (y_{2i}) = 
		 	\half (\, 3^{n-i+1}  - 3^{n+1}\, )
			&&\qquad \text{ for $1\le i \le n$} \\
	\eta^{-} (x_{2i+1}) &= \eta^{-} (x_{2i})  - 2 \cdot 3^{n-i} 
	=  	\half (\, - 3^{n-i}	-  3^{n+1} \,)
	&&\qquad \text{ for $1\le i < n$} 
\end{align*}
Of course, $\eta^{-} (x_0) = 0$ , $\eta^{-} (y_0) = 0$, and $\eta^{-} (x_1) = -2 \cdot 3^{n}$. Now observe that in the Dijkstra phase, 
the order of relaxations does not matter: 
every non-negative edge is succeeded by only negative edges, so potentials only get propagated one edge forward, which is
equivalent to relaxing the non-negative 
edges once in an arbitrary order. The only affected vertices are the even-indexed ones $x_{2i}, y_{2i}$ for $1 \le i \le n$; in particular we have
\begin{align*}
	\eta (x_{2i}) &= \min\braced{ \eta^{-} (x_{2i}) , \eta^{-} (x_{2i-1}) + 2 \cdot 3^{n-i}}  
	\\
	&= \min\braced{\half (3^{n-i+1}  - 3^{n+1}),  	
				\half (- 3^{n-i+1}	-  3^{n+1}) + 2\cdot 3^{n-i}} 
	\\
	&= \min\braced{\half (3^{n-i+1}  - 3^{n+1}),  	
				\half (3^{n-i}	-  3^{n+1})} 
	\\
	&= \half (3^{n-i}	-  3^{n+1}) 
	\\
	\eta(y_{2i} ) &= \min\braced{ \eta^{-} (x_{2i}) , \eta^{-} (x_{2i-1}) +  3^{n-i}}  
	\\
	&= \min\braced{\half (3^{n-i+1}  - 3^{n+1}), 
					\half (- 3^{n-i+1}	-  3^{n+1})+  3^{n-i}}  
	\\
	&= \min\braced{\half (3^{n-i+1}  - 3^{n+1}), 
					\half (- 3^{n-i}	-  3^{n+1})}
	\\
	&= \half (- 3^{n-i}	-  3^{n+1})
\end{align*}
(That is, the paths going through $x_{2i-1}$ are strictly shorter.) 
For the remaining vertices $\eta(v)= \eta^{-} (v)$. By straightforward calculation (see below), one can show that for $1 \le i < n$:
\begin{align*}
	\ell_\eta(x_{2i} , x_{2i+1}) &= \ell(y_{2i} , y_{2i+1}) \\
	\ell_\eta (y_{2i}, y_{2i+1}) &= \ell(x_{2i}, x_{2i+1}) \\
	\ell_\eta (x_{2i+1}, x_{2i+2}) &= \ell (y_{2i+1}, y_{2i+2}) \\
	\ell_\eta (y_{2i+1}, y_{2i+2}) &= \ell (x_{2i+1}, x_{2i+2}) \\
	\ell_\eta (x_{2i+1}, y_{2i+2}) &= \ell (y_{2i+1}, x_{2i+2}) \\
	\ell_\eta (y_{2i+1}, x_{2i+2}) &= \ell (x_{2i+1}, y_{2i+2}) 
\end{align*}
For the case $i=0$, all values $\ell_\eta (x_0, x_1),\ell_\eta(y_0, y_1), \ell_\eta(x_1, y_2), \ell_\eta(x_1, x_2)$ are zero, 
while $\ell_\eta(y_1, x_2) = 3^{n}$ and $\ell_\eta (y_1, y_2) = 2\cdot 3^{n}$. Because these six edges are non-negative, the Elmasry potentials for $x_0, x_1, x_2, y_0, y_1$ will be zero 
(realized by empty paths)
for all later iterations of Elmasry's algorithm. Furthermore, for $j \ge 2$ the shortest nbp-paths ending at $x_j$ or $y_j$ in later iterations never start at $x_0, x_1, y_0, y_1$. (Or more precisely, there exists a shortest nbp-path that does not begin at these vertices.)

Consequently, the later Elmasry potentials depend only on the sub-graph induced by $V_n \setminus\braced{x_0, x_1, y_0, y_1}$ with weights $\ell_\eta$. But by the above calculations, this sub-graph is identical to $G_{n-1}$, with the roles of $x_j$ and $y_j$ swapped. By the induction hypothesis, the algorithm must execute $n-1$ more iterations, finishing the proof.

For completeness, here are the calculations for the reduced weights: for $1\le i < n$ we have
\begin{align*}
\ell_\eta(x_{2i} , x_{2i+1}) &= \ell(x_{2i} , x_{2i+1}) + \eta(x_{2i}) - \eta(x_{2i+1}) 
\\
&= -2 \cdot 3^{n-i}	 + \half (3^{n-i}	-  3^{n+1}) - \half (-3^{n-i}	-  3^{n+1}) = -3^{n-i}
\\
&= \ell(y_{2i}, y_{2i+1}) 
\end{align*}
\begin{align*}
\ell_\eta (y_{2i}, y_{2i+1}) &=  \ell (y_{2i}, y_{2i+1}) + \eta(y_{2i}) - \eta(y_{2i+1})
\\
&=  -3^{n-i} + \half (-3^{n-i}	-  3^{n+1}) - \half (3^{n-i}	-  3^{n+1}) = -2\cdot 3^{n-i}
 \\
&= \ell(x_{2i}, x_{2i+1}) 
\end{align*}
\begin{align*}
\ell_\eta (x_{2i+1}, x_{2i+2}) &=  \ell (x_{2i+1}, x_{2i+2}) + \eta(x_{2i+1}) - \eta(x_{2i+2})
\\
&=  2\cdot 3^{n-i-1} + \half (-3^{n-i}	-  3^{n+1}) - \half (3^{n-i-1}	-  3^{n+1}) = 0
\\
&= \ell(y_{2i+1}, y_{2i+2}) 
\end{align*}
\begin{align*}
\ell_\eta (y_{2i+1}, y_{2i+2}) &=  \ell (y_{2i+1}, y_{2i+2}) + \eta(y_{2i+1}) - \eta(y_{2i+2})
\\
&=  0 + \half (3^{n-i}	-  3^{n+1}) - \half (-3^{n-i-1}	-  3^{n+1}) =  2  \cdot 3^{n-i-1}
\\
&= \ell (x_{2i+1}, x_{2i+2}) 
\end{align*}
\begin{align*}
\ell_\eta (x_{2i+1}, y_{2i+2}) &=  \ell (x_{2i+1}, y_{2i+2}) + \eta(x_{2i+1}) - \eta(y_{2i+2})
\\
&=   3^{n-i-1} + \half (-3^{n-i}	-  3^{n+1}) - \half (-3^{n-i-1}	-  3^{n+1}) = 0
\\
&= \ell (y_{2i+1}, x_{2i+2}) 
\end{align*}
\begin{align*}
\ell_\eta (y_{2i+1}, x_{2i+2}) &=  \ell (y_{2i+1}, y_{2i+2}) + \eta(y_{2i+1}) - \eta(x_{2i+2})
\\
&=  0 + \half (3^{n-i}	-  3^{n+1}) - \half (3^{n-i-1}	-  3^{n+1}) = 3^{n-i-1} 
\\
&=\ell(x_{2i+1}, y_{2i+2})
\end{align*}


\section{Analysis for Paths}
\label{sec: analysis for paths}

In this section we analyze the running time of Elmasry's algorithm if the graph is just a path.
We show that in this case the number of iterations is $O(\log n)$, and show an example
where $\Omega(\log n)$ iterations are necessary.


\subsection{Upper Bound for Paths}
\label{subsec: upper bound for paths}
Here we show that on a path of length $n$ the number of iterations of Elmasry's algorithm
is $O(\log n)$. Since $G$ is a path, its subgraph $G^-$ of non-positive edges consists of disjoint segments.
The rough idea is to show that these segments of $G^-$ get merged from one
phase to next, reducing the number of segments geometrically.
In the more formal argument below, we'll show that we can as well assume that
there are no zero weights and that the negative and positive weights alternate,
and using this simplification we then show that the number of negative weights
will at least halve in each iteration.

We start with some intuition. Let $G$ be the path $v_1\edgeto v_2 \edgeto ... \edgeto v_n$. Let $\weight_i$ be the weight of an edge $v_i \edgeto v_{i+1}$, 
for each $i$.
For now, suppose that all weights on this path are non-zero. 
For $p < q$, let $\weight_{p,q} = \sum_{s = p}^{q-1}\weight_s$ be the sum of the weights on the path segment from $v_p$ to $v_q$.
Consider two negative edges $v_i \edgeto v_{i+1}$ and $v_j \edgeto v_{j+1}$ with $j > i+1$, such that
edge $v_{i-1}\edgeto v_i$ is positive (or $i=1$), and all edges in-between $v_{i+1}$ and $v_j$ are positive (there is least one).
Elmasry's algorithm will propagate the potential from $v_i$ along the path to $v_j$,
so for $q = i+1,...,j$, Elmasry's potential for $v_q$ will be $\eta(v_q) = \min (\weight_{i,q},0)$.
Therefore, what Elmasry's algorithm will do in this iteration is this:
\begin{itemize}
\item If $\weight_{i,j}\ge 0$, let $q \le j$ be largest for which $\weight_{i,q} \le 0$.
All reduced weights between $v_i$ and $v_q$ will become $0$, all reduced weights between $v_q$ and $v_j$ will remain positive (because the
potentials along this segment will be $0$), and $v_j \edgeto v_{j+1}$ will have reduced weight $0$. 
Furthermore, the reduced weight of the $v_i$-to-$v_j$ path is still $\ell_{i,j}$.
\item If $\weight_{i,j} < 0$, then
all reduced weights between $v_i$ and $v_j$ will become $0$, and the reduced weight of $v_j\edgeto v_{j+1}$ will remain negative 
(in particular, it has reduced weight $\ell_{i,j}$).
\end{itemize}

$G$ may have several negative edges in a row. This does not affect the above observation, as the
only difference is that we need to consider above the block starting at a negative edge $v_i \edgeto v_{i+1}$,
consisting of some negative edges and followed by some positive edges, and ending right before the first
negative edge $v_j \edgeto v_{j+1}$ after these positive edges.

This can be further generalized to allow zero-weight edges. These edges do not affect the above
behavior: they can be simply eliminated from $G$, as their weights will remain zero, and the weights
of other edges will change as if these zero edges were not present.

Consider one iteration. We will express our arguments in terms of the weights of the edges in $G$.
Let the weight sequence be $P = \weight_1, \weight_2, ..., \weight_n$. 
As explained above, zero-weight edges can be removed from $P$, and any two consecutive weights with the
same sign can be combined into one. Call thus obtained sequence $P'$ a \emph{contraction} of $P$.
For simplicity, we can also assume that $n$ is even and $\ell_n$ is positive. 

So instead of $P$ we can consider the contraction $P'$ of $P$, where $P' = \weight_1, \weight_2, ..., \weight_n$,
$n$ is even, all odd-indexed weights are negative and all even-indexed weights are positive.
Call an odd index $i$ \emph{terminal} if $\weight_i + \weight_{i+1} > 0$.
For two indices $i < j$,
where $i$ is odd and $j$ is even,
 an interval $[i,j]$ is called an \emph{neg-segment}
if $i-2$ is terminal (or $i=1$), $j-1$ is terminal, and $i,...,j-3$ are non-terminal.

As a result of applying one iteration of Elmasry's algorithm, the weight sequence in a neg-segment $[i,j]$
will change as follows:
\begin{equation*}
\begin{array}{ccccccccccc}
\textrm{before} & 
\weight_i & \weight_{i+1} & \weight_{i+2} & \weight_{i+3} & \weight_{i+4} & ....  & \weight_{j-3} & \weight_{j-2} & \weight_{j-1} & \weight_j 
\\
\textrm{after} &
0 & 0 & \weight'_{i+2} & 0 & \weight'_{i+4} &  .... & \weight'_{j-3} & 0 & \weight'_{j-1} & \weight'_j 
\end{array}
\end{equation*}
where $\weight'_{i+2}, \weight'_{i+4}, ..., \weight'_{j-1} \le 0$ and $\weight'_j > 0$.
Therefore:
\begin{itemize}

\item If $j = i+1$, then after the iteration the interval $[i,j]$ will not have a negative weight.

\item If $j \ge i+3$, then after the iteration in the interval $[i,j]$ only one positive weight will be left (namely $\weight'_{j}$).

\end{itemize}
Let $Q$ be the sequence after the iteration and $Q'$ be its contraction.
From these properties, if $j = i+1$ then no negative weight will
remain from $[i,j]$, and if $j \ge i+3$ then at most one negative weight will remain (and there were at least two).
This implies that
the number of negative weights decreases by a factor of at least $2$
from $P'$ to $Q'$. So the number of iterations will be $O(\log n)$.


\subsection{Lower Bound for Paths}
\label{subsec: lower bound for paths}

We now show an example where $G$ is a path of length $n$ on which Elmasry's algorithm will
make $\Omega(\log n)$ iterations before all weights become non-negative.
As in the previous section,
we will identify the path by its weight sequence $\weight_1, \weight_2, ..., \weight_n$.
The construction is recursive.

For $n=2$, the sequence is $-1,1$. In this case, the process will end with non-negative weights
after $1$ iteration.

Suppose that $P = \weight_1,\weight_2,...,\weight_n$ is a sequence that requires $s$ iterations,
where $n$ is even,
and has the property that all odd-indexed weights are negative and all even-indexed weights are positive.
We show how to modify this sequence, doubling its length, so that $s+1$ iterations will be required.

First, as we noted earlier, adding edges with zero weights does not affect the changes of the
non-zero weights. So the sequence
\begin{equation*}
		P' = 0,0,\weight_1,\weight_2,0,0,\weight_3,\weight_4,...,0,0,\weight_{n-1},\weight_n
\end{equation*}
will also require $s$ iterations. Now, let path $Q$ be obtained from $P'$ as follows:
for each even $i$, replace the 4-tuple $0,0,\weight_{i-1},\weight_i$ by  $\weight_{i-1}-1,1,-1,\weight_i+1$.

Then, during the first iteration on $Q$, Elmasry's algorithm will convert the first 4-tuple $\weight_{1}-1,1,-1,\weight_2+1$
into $0,0,\weight_1,\weight_2$. Thus the negative values will not spill to the second
4-tuple, so the second 4-tuple $\weight_{3}-1,1,-1,\weight_4+1$ will be
converted to $0,0,\weight_3,\weight_4$, and so on. Overall, $Q$ will be converted into $P'$.
This implies that $Q$ requires $s+1$ steps, as claimed.
Further, in $Q$ there are no zero values, and positive and negative values alternate.
So this construction can be applied again to $Q$.


\bibliographystyle{alpha}
\bibliography{./shortest_paths.bib}

\end{document}